\documentclass[ french, 12pt]{article}
\usepackage{amssymb}
\usepackage{graphicx}
\usepackage{url}

\begin{document}

\begin{center}

{\Large On Integrability and Exact Solvability in Deterministic and Stochastic Laplacian Growth}

\vspace{5mm}

{\large Igor Loutsenko and Oksana Yermolayeva}

\vspace{5mm}

{\small Laboratoire de Physique Math\'ematique,\\
CRM, Universit\'e de Montr\'eal\\[1mm]}

\vspace{5mm}

\end{center}

\begin{abstract}
We review applications of theory of classical and quantum integrable systems
to the free-boundary problems of  fluid mechanics as well
as to corresponding problems of statistical mechanics.
We also review important exact results obtained in the theory of multi-fractal spectra of the stochastic models related to the Laplacian growth: Schramm-Loewner and Levy-Loewner evolutions.
\end{abstract}

\begin{section}{Introduction and Outline}

\label{Introduction}

    The theory of diffusive growth processes unites
the mathematical modeling of numerous non-equilibrium phenomena
that involve interaction of several phases of matter,
such as oil and water in the Hele-Shaw cell, solid and
liquid states in  crystal growth, different electronic states in magnetic fields and many others. This theory deals with an important class of pattern formation of moving fronts which occurs when diffusion rather than convection dominates the transport. It studies models of essentially two types:

The models of the first, ``hydrodynamic", type are deterministic
free-boundary problems. The simplest class of these is usually
unified by the name of Laplacian Growth, also known as the Hele-Shaw
problem, which refers to dynamics of a moving front (``boundary" or
``interface") between two distinct phases in two dimensions driven by
a harmonic scalar field that is a potential for the growth velocity field
(see e.g. \cite{BKLST}, \cite{VE}, and \cite{GTV} that contains more up to date results). Such models
describe for example fingering viscous flows in fluid
mechanics.

The models of the second type are stochastic versions of the above
deterministic problems introduced to describe processes where interfaces form natural fractal patterns. The most famous representative of the second type,
the Diffusion Limited Aggregation or DLA (for introduction see e.g. \cite{H}, \cite{LYZ}), describes, for instance, formation of a
cluster of particles on two dimensional lattice by aggregation of random walkers released from infinity: No more than one
particle can occupy a lattice site. The particles stick one by one
to the cluster. Then it follows that the probability for a particle to occupy a given, next to the cluster,
site depends on the value of a lattice
harmonic field at neighboring sites. The field vanishes on the cluster and
has a prescribed asymptotic behavior far from the boundary.

The main targets of
research in Laplacian growth is to predict  dynamics of
the boundary and/or formation of patterns between different phases as well as to understand principles that unite various growth models in universality classes (for instance, Hele-Shaw problem and DLA are claimed to be in the same universality class \cite{MPST}). The indirect goals here aim to study mathematical structures behind the processes and to establish links with adjacent disciplines, like potential theory and quadrature domains, integrable systems and exactly-solvable models of statistical mechanics as well as specific fundamental problems of mathematical physics.

In this work we review exact results obtained in the Laplacian growth and related models. In particular, we consider application of theory of integrable systems to the models of the ``hydrodynamic" type and reveal the established links with specific fundamental problems of mathematical physics (Sections \ref{Toda} and \ref{Elliptic}). We also review attempts of integrable discretizations of hydrodynamic models in the framework of equilibrium statistical mechanics through theory of random matrices and infinite soliton solutions of integrable hierarchies (Section \ref{Coulomb}). The last part of the review is devoted to exact results obtained in the theory of  Schram-Loewner and Levy-Loewner evolutions (Section \ref{LLE}).

\end{section}

\begin{section}{Deterministic and Stochastic Laplacian Growth}
\label{LG_DLA}

A zero surface tension Hele-Shaw free-boundary
problem (see e.g. \cite{BKLST}, \cite{GTV}, \cite{VE}) is a deterministic version of the Laplacian Growth. It describes time evolution of a
domain $\Omega=\Omega(t)$ in the $z=x+iy$ plane, when the boundary
$\partial\Omega$ of the domain is driven by the gradient of a scalar
field $P=P(x,y,t)$, often referred as ``pressure" (see Fig. \ref{continuous}):
The normal velocity of the boundary $\partial\Omega$ is proportional to the gradient of the pressure
\begin{equation}
v_n=-\partial_n P ,
\label{velocity}
\end{equation}
where $v_n$ stands for normal velocity of the boundary and $\partial_n$ is the normal derivative at the boundary.

The field is harmonic, except for several singular points (``sources" and ``sinks"), in
the exterior or respectively the interior of $\Omega$:
$$
\Delta P(x,y,t)=\sum_iq_i(t)\delta(x-x_i)\delta(y-y_i),
$$
The corresponding Hele-Shaw problems are referred as exterior or interior. In the most known case of the exterior Laplacian growth driven by a single sink at infinity the above equation takes the form
\begin{equation}
\Delta P=0, \quad P \to \frac{-1}{4\pi}\log(x^2+y^2) \,\, {\rm as} \,\, z\to\infty.
\label{LG}
\end{equation}

The field $P$ vanishes at the domain boundary
(``interface")
\begin{equation}
P(\partial\Omega(t),t)=0.
\label{boundary}
\end{equation}
The domain evolution is defined completely by its initial shape and behavior of singularities $q_i(t)$.
The well-posedness and
stability of the problem depend strongly on types of the singularities.

\begin{figure}
\centering
\includegraphics[width=150mm]{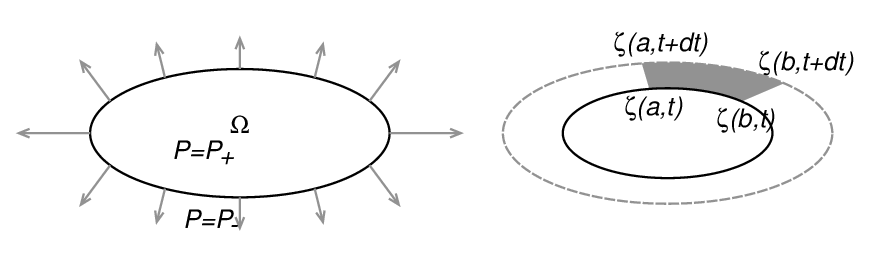}
\caption{Problem setting (left) and the domain area increment along
the boundary segment $a<l<b$ during the time interval $(t,t+dt)$
(right). For an exterior problem $P_+=0$, and $P_-=0$ for an interior problem. } \label{continuous}
\end{figure}

For instance, the exterior ``sink"-driven Hele-Shaw problem, i.e.
the problem of expanding of simply-connected domain $\Omega$ whose
boundary is driven by the Green's function of the exterior of $\Omega$, is
linearly unstable (so-called Saffman-Taylor instability \cite{ST}) and ill-posed for almost any initial conditions (for details see e.g. \cite{JS} or section 1.4.3 of \cite{GTV}) \footnote{The problem can be  naturally regularized by taking into account the surface tension of an interface (see e.g. \cite{ES} or section 1.4.4 of \cite{GTV}). Unfortunately, this generalization is no longer integrable.}.
In the contrast, the interior problem for expanding $\Omega$ (which is
a ``source" driven) is linearly stable and well-posed.

One may think of the above free boundary problem as that of dynamics of an ideal
conducting contour $\partial\Omega$ in two-dimensional electric
field. In the case of the exterior Laplacian growth driven by the logarithmic singularity at infinity (\ref{velocity},\ref{LG},\ref{boundary}),
a unit charge is distributed with linear density
$\sigma(l,t)$ along the contour. The contour is ``an ideal
conductor", that means that the potential (at fixed $t$) is
constant along $\partial\Omega$. Here $l$ stands for natural
parameter along the contour $\left\{\partial\Omega: z=\zeta(l,t),
|d\zeta(l)|=dl\right\}$ and $\oint_{\partial\Omega}\sigma(l,t)dl=1$.
The harmonic field $P(x,y,t)$ is (modulo coordinate independent
function of $t$) the electrostatic potential created by the above
charges
\begin{equation}
P=\frac{1}{2\pi}\oint_{\partial\Omega}\sigma(l,t)\log|z-\zeta(l,t)|dl .
\label{contour}
\end{equation}
The gradient of the potential is the electric vector field, that has
a jump of magnitude $\sigma(l)$ across the boundary. Therefore, from
(\ref{velocity}) it follows that the normal velocity of the contour is
\begin{equation}
v_n(l,t)=\sigma(l,t) .
\label{VSigma}
\end{equation}
Equations (\ref{contour}), (\ref{VSigma}) reduce the free-boundary problem on the
plane to the dynamical problem on the contour only. It is easy to see that
the linear density $\sigma$ is non-negative, and the interior domain
$\Omega$ expands in time.

The domain area increment $d{\cal P}_{(a,b)}$ along the interface segment
$a<l<b$ during the time $dt$ (see Fig. \ref{continuous}) is proportional to the electric charge $q_{(a,b)}$ concentrated on
this segment
\begin{equation}
\frac{d{\cal P}_{(a,b)}}{d
t}=\int_{a}^{b}v_n(l)dl=\int_{a}^{b}\sigma(l)dl=q_{(a,b)} .
\label{probability}
\end{equation}
This fact suggests to consider the boundary charge as
the cluster increment probability in the discrete analogue of the deterministic Laplacian growth called diffusion-limited aggregation (DLA). This is a stochastic growth process that is yet analogous in its derivation to the above model.

\begin{figure}
\centering
\includegraphics[width=115mm]{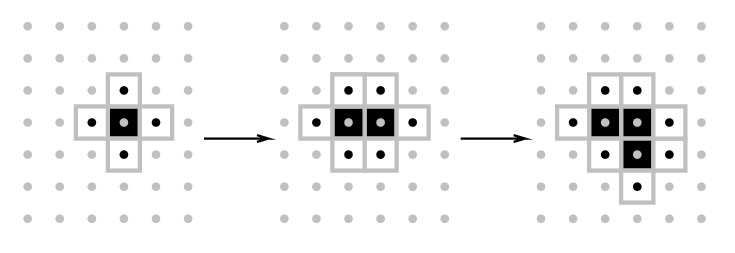}
\caption{An example of the cluster growth. Black squares
(``particles") belong to the cluster $\Omega$, white squares with
gray sides form the cluster boundary $\partial\Omega$. Black dots
denote charged (boundary) sites and gray dots are uncharged lattice
sites.}\label{explaincluster}
\end{figure}

Let us now take a square lattice, paint elementary square cells in
white color and label centers of squares by a pair of integers
$(m,n)$ that are discrete coordinates along the $x$ and $y$
directions. Consider the following stochastic process: At the first
step, the $(0,0)$ square is painted in black. At the next step we
paint in black one of four sites that are next-neighbors of this
smallest cluster etc. (see Figure \ref{explaincluster}). This
procedure is continued by coloring, at each step, one of randomly
chosen white next neighbors of the cluster (i.e. adding randomly a
``particle" to the cluster) with the probability described below.

\begin{figure}
\centering
\includegraphics[width=111mm]{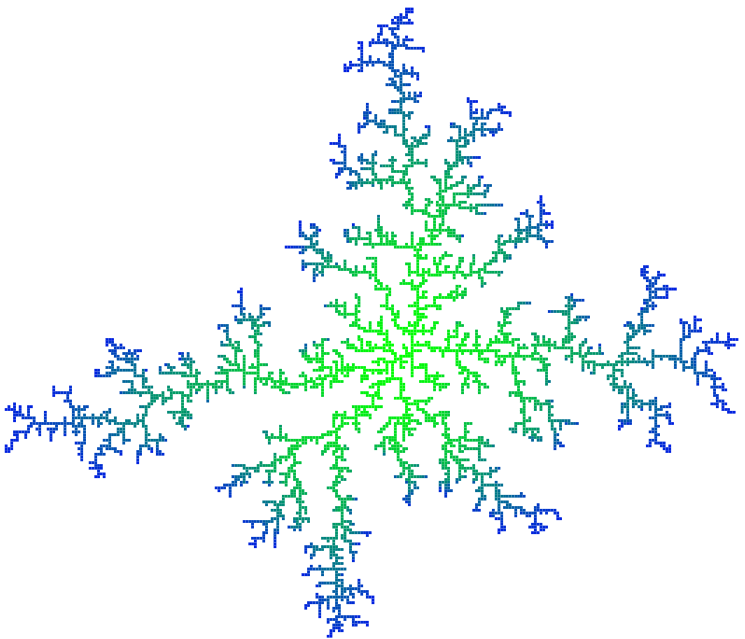}
\caption{Example of the growing DLA cluster. To visualize
the time dynamics, we prescribe colors to cluster particles. The color ranges
from green to blue, depending on time (or step) at which the particle joined the
cluster (green for the initial step, blue for the final step). }
\label{DLA}
\end{figure}

We use the same notation $\Omega$ for the cluster as we used for a
continuous domain. The cluster boundary, which consists of all white
next-neighbors of the cluster, is denoted, by analogy with
continuous case, by $\partial\Omega$.

At each step of the process, we can define a lattice function
$P_{m,n}$ which satisfies the difference Laplace equation
$$
P_{n-1,m}+P_{n+1,m}+P_{n,m-1}+P_{n,m+1}-4P_{n,m}=0,
$$
$$
P_{n,m}\to\frac{-1}{4\pi}\log(m^2+n^2), \quad
m^2+n^2\to\infty,
$$
and vanishes at the boundary
$$
P_{n,m}=0, \quad (n,m)\in\partial\Omega. \label{lattice_boundary}
$$
The above equations are lattice analogues
of (\ref{velocity}, \ref{LG}, \ref{boundary}).

As in continuous case, the harmonic field $P_{n,m}$ can be also
expressed in terms of charges, that are now sitting at the boundary
sites (see Fig. \ref{explaincluster}).

It is easy to see that the boundary charges are nonnegative $q_{i,j}\ge
0$ and
$$
\sum_{(i,j)\in\partial\Omega}q_{i,j}=1.
$$
By analogy with continuous case (c.f. (\ref{probability})) one
can consider $q_{i,j}$ as a probability of the $(i,j)$ square of the
boundary $\partial\Omega$ to join the cluster $\Omega$.

The above discretization of the Laplacian growth is a ``site" version of the Diffusion Limited Aggregation or DLA, a stochastic process of cluster growth by consecutive attachment to its boundary random walkers released from infinity \cite{H}, \cite{LYZ}. An example of such a cluster consisting of several thousand square particles is shown on Figure \ref{DLA}.

The ill posedness of the continuous Laplacian growth driven by sinks exhibits as fractal growth in its stochastic version. Experimental results suggest that the Hele-Show problem and DLA are in the same universality class in the sense that they have the same multi-fractal spectrum \cite{MPST}. Theoretical demonstration of this fact is an important unsolved problem.

\end{section}

\begin{section}{Laplacian Growth as Solution of Hierarchies of Integrable PDEs}
\label{Toda}

The deterministic Laplacian growth (the Hele-Shaw problem) possesses a remarkable integrable structure: It was found in 1970 by S. Richardson \cite{R} that, for this problem, there exists an infinite set of integrals of motion
$$
I[h]=\int_{\Omega(t)} h(x,y) dx dy, \quad dI/dt=0
$$
where $h(x,y)$ is any regular in $\Omega$ harmonic function that vanishes at sources/sinks.

It turned out that given initial form of domain at $t=0$, its final form does not depend on a history of sources $q_i(t)$ but rather on total fluxes $Q_i=\int_0^t q_i(t')dt'$ produced by the time $t$. In other words, one observes ``commutativity" of fluid flows driven by different sources.

In the case of the exterior Laplacian growth driven by a single source/sink at infinity one can take harmonic moments of the exterior domain $\mathbb C\backslash\Omega$ as invariants of motion
\begin{equation}
I_k=\int_{\mathbb C\backslash\Omega(t)} z^{-k} dx dy, \quad {\bar I}_k=\int_{\mathbb C\backslash\Omega(t)} \bar z^{-k} dx dy, \quad k=1,2, \dots
\label{HI}
\end{equation}
More precise notion of integrability of Laplacian growth in simply connected domains was given in context of the dispersionless 2dToda hierarchies in works \cite{MWZ}, \cite{HLY}, and in context of the Whitham hierarchy for multiply connected domains \cite{KMWZ}.

\begin{figure}
\centering
\includegraphics[width=105mm]{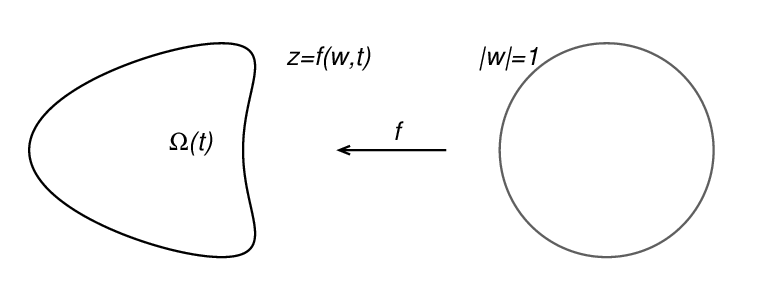}
\label{Riemann}
\caption{Time dependent conformal mapping $z=f(w,t)$ from the unit circle in "mathematical" $w$-plane to the domain boundary in ``physical" $z$-plane. Depending on the problem, exterior/interior of the circle is mapped analytically to the exterior/interior of $\partial\Omega(t)$ respectively.}
\end{figure}

For the simply connected case, considered here, it is convenient to use a time-dependent conformal mapping for description of the domain evolution: By Riemann's theorem one can analytically map an interior or exterior of the unit circle in the ``mathematical" $w$-plane onto respectively interior $\Omega$ or exterior $\mathbb C\backslash\Omega$ of the ``physical" plane, so the domain boundary $\partial\Omega: z=f(e^{i\theta},t)$ is the image of the circle $|w|=1$ (see Figure \ref{Riemann}).

Considering an exterior problem one can write down
\begin{equation}
f(w,t)=r(t)w+\sum_{i=0}^\infty u_k(t)w^{-k}
\label{conformalLG}
\end{equation}
and it can be shown that a solution of the exterior Hele-Shaw problem satisfies the Polubarinova -Kochina equation \cite{PK}, which modern form is \cite{MWZ}
\begin{equation}
\{f(w,t), \bar f(1/w,t)\}=1 .
\label{string}
\end{equation}
Here ``Poisson brackets" $\{,\}$ are defined as
\begin{equation}
\{f(w,t), g(w,t)\}:=w\frac{\partial f}{\partial w}\frac{\partial g}{\partial t}-w\frac{\partial g}{\partial w}\frac{\partial f}{\partial t}.
\label{quasiclassic}
\end{equation}
It has been observed in \cite{MWZ} that, when considered as a quasi-classical limit of matrix commutator in Lax representation, the Laplacian growth (the Polubarinova-Kochina) equation (\ref{string}) coincides with so-called ``string" constraint to solutions of the dispersionless 2Toda hierarchy (2dToda), so that the domain deformation due to the change of different harmonic moments corresponds to evolutions wrt different hierarchy times.

The 2Toda hierarchy is an integrable infinite-dimensional dynamical system which can be constructed in the following way. Define two infinite-dimensional Lax matrices: let the first one have zero entries above the first upper diagonal and the second one be with zeros below the first lower diagonal:
$$
L=r(T)W+\sum_{i=0}^{\infty}u_i(T)W^{-i}, \quad \bar L=W^{-1}r(T)+\sum_{i=0}^{\infty}\bar u_i(T)W^{i}.
$$
Here $i\in \mathbb Z$ and $T \in \mathbb Z$ are matrix indexes: $i$ stands for the number of the matrix diagonal and $T$ labels elements along the diagonal. $W$ is the $T$-shift operator $W[g](T)=g(T+1)$, i.e. an infinite-dimensional matrix with units on the first upper diagonal and zeros elsewhere $W_{ij}=\delta_{i,j+1}$, $i,j\in Z$.

The 2Toda system describes a set of continuous deformations of Lax matrix elements $r(T), u_i(T), \bar u_i(T)$ through the following equations
$$
\frac{\partial L}{\partial t_k}=[H_k, L], \quad \frac{\partial L}{\partial \bar t_k}=[\bar H_k, L],
$$
$$
\frac{\partial \bar L}{\partial t_k}=[H_k, \bar L], \quad \frac{\partial \bar L}{\partial \bar t_k}=[\bar H_k, \bar L],
$$
where $t_k, \bar t_k, k=1,2,\dots $ are corresponding continuous deformation parameters and $[,]$ stands for the matrix commutator.

In the above equations \footnote{Here, we used the ``symmetric gauge" format of 2dToda equations.}, the evolution operators $H_k, \bar H_k$, $k=1,2, \dots$ are defined as follows
$$
H_k=(L^k)_++\frac{1}{2}(L^k)_0, \quad \bar H_k=(\bar L^k)_-+\frac{1}{2}(\bar L^k)_0 ,
$$
where $()_\pm$ denotes projection to upper/lower ``triangular" part of the matrix and $()_0$ denotes its main diagonal part, i.e. for a matrix $A=\sum_{k=-\infty}^\infty a_kW^k$
$$
A_+:=\sum_{k=1}^\infty a_kW^k, \quad A_-:=\sum_{k=-\infty}^{-1} a_kW^k, \quad A_0:=a_0 .
$$
The 2Toda hierarchy is a set of equations describing deformations of a one-dimensional lattice with sites labeled by an integer $t$: The dispersionless (2dToda) hierarchy is obtained as the long-wavelength (or quasiclassic) limit of the 2Toda hierarchy:
$$
r(T)-r(T+1) \to 0, \quad u_i(T)-u_i(T+1) \to 0, \quad \bar u_i(T)-\bar u_i(T+1) \to 0 .
$$

In this limit, the shift operator $W$ is replaced with a continuous variable $w$ , and the lattice index $T$, after a proper rescaling, also becomes a continuous variable $t$. The Lax matrix $L$ becomes a Lax function $f(w,t)$ of the form (\ref{conformalLG}). The matrix commutators transforms into the brackets (\ref{quasiclassic}): $[,]\to\{,\}$. The evolution operators become functions obtained by projections of $f(w,t)^k, \bar f(1/w,t)^k$ to the positive/negative power and constant parts of the corresponding Laurent series in $w$.

Interpreting the conformal mapping in Laplacian growth (\ref{conformalLG}) as a Lax function, one can see that a Laplacian growth equation (\ref{string}) is preserved by all (properly rescaled) $t_k, \bar t_k$-deformations and defines reductions of the dispersionless hierarchy. Therefore, 2dToda hierarchy governs symmetries of the Laplacian growth. In this picture, the hierarchy ``times"  can be interpreted as the harmonic moments  (\ref{HI}) of the domain, i.e. (up to constant shift)
$$
t=\int_\Omega dxdy, \quad t_k=I_k, \quad \bar t_k=\bar I_k , \quad k=1,2, \dots
$$
It is important to note that $t_k, \bar t_k$-deformations do not include evolution in physical time $t$ of the Laplacian growth, as $t$ is not an evolution parameter of the 2dToda, but rather a ``continuous lattice" coordinate of this hierarchy. Deformation by $t_k, \bar t_k$-flow only changes $k$th harmonic moment of the domain $\Omega(t)$, but not its area. This kind of motion can be considered as an evolution of a domain driven by pure $k+1$-pole source, i.e. by a singularity of multiplicity $k+1>1$.

Since the above connection with hierarchies of integrable PDEs gives us a Lax representation of the Laplacian growth with the ``physical" time $t$ playing role of the lattice index, rather than evolutional parameter of 2dToda, a further work on finding Hamiltonian structure of the Laplacian growth is needed: Establishing a Hamiltonian or variational formulations of the Laplacian growth is one of the central problems in interface dynamics that could provide proper generalizations of the boundary dynamics in the framework of statistical mechanics. The hamiltonian structure of physically important finite-dimensional reductions, describing solutions of the exterior Laplacian growth ($N$-finger solutions, see e.g. \cite{PM}), that do not blow up in zero time, has been studied in e.g. in work \cite{HLY}.

\end{section}

\begin{section}{Normal Random Matrix and Coulomb Gas Interpretation of Interface Dynamics}
\label{Coulomb}

An alternative to the Lax representation of an integrable hierarchy is presented by its Hirota form, where instead of Lax equations one solves a system of bi-linear differential/difference equations for the $\tau$-function (see e.g. \cite{MJD}). In our case $\tau=\tau(T; t_1,\bar t_1, t_2, \bar t_2, \dots)$ is a function that contains all information about $r, u_i, \bar u_i$.

It is known, that $\tau$-functions of special classes of solutions of integrable hierarchies are partition functions of various (depending on a choice of hierarchy) random matrix ensembles. In particular, the $\tau$-function of 2Toda hierarchy restricted by the ``string equation" $[L, \bar L]=1$ (with latter being a dispersive analog of the Laplacian growth equation (\ref{string})), is nothing but a partition function of the normal random matrix model \cite{GMMMO, KKMWZ, Z}
$$
\tau(T; t_1, \bar t_1, t_2, \bar t_2, \dots)=\int_{[M, M^+]=0}e^{-{\rm Tr}\left(\frac{1}{\hbar} M^+M+\sum_{k>0}t_kM^k+\bar t_k(M^+)^k\right)}  dMdM^+ ,
$$
where the integral is taken over all normal (i.e. $M$: $[M, M^+]=0$, where $M^+$ is the hermitian transpose of $M$) matrices of size $T\times T$ and $\hbar$ being a free parameter. The dispersionless limit corresponds to a double-scaling limit for large matrices, a such that $T\to \infty$ and $\hbar\to 0$, while $t=\hbar T$ remains finite. Deformation variables in this limit are also need to be rescaled $t_k\to t_k/\hbar $.

Since, in the dispersionless limit, the above $\tau$-function describes both the Laplacian growth and normal random matrices, it is natural
 to ask how the eigenvalue distribution of these random matrices is related with domain shape $\Omega(t)$ in the Hele-Shaw problem?

A normal matrix can be represented as $M={\cal U}\Lambda {\cal U}^+$, where ${\cal U}$ is a unitary matrix and $\Lambda={\rm diag}(z_1, z_2, \dots z_T)$ is a diagonal matrix consisting of complex eigenvalues of $M$. By a standard procedure (see e.g. review \cite{MPT}), one can integrate out ``angular" ${\cal U}$-variables and write down the eigenvalue probability distribution:
$$
d{\cal P}=\frac{1}{\cal Z} e^{-E} \prod_{i=1}^T dx_idy_i, \quad {\cal Z}=\int  e^{-E} \prod_{i=1}^T dx_idy_i,
$$
where $E$ is an electrostatic energy of the two-dimensional Coulomb gas:
\begin{equation}
E=-2\sum_{1 \le i<j \le T} \log|z_i-z_j|+\frac{1}{\hbar}\sum_{i=1}^T (z_i\bar z_i+V(z_i, \bar z_i)), \quad V(z, \bar z)=\sum_{k>0} (t_k z^k+ \bar t_k \bar z^k).
\label{GEnergy}
\end{equation}
The above probability distribution corresponds to the particular model of random normal matrices that is reduced to statistical mechanics of $T$ mutually repelling positive Coulomb charges in the confining potential $z\bar z/\hbar$ produced by uniformly charged negative background on the plane. In addition, the harmonic potential $V(z,\bar z)/\hbar$, $\Delta V=0$, acting on each charge, is superposed on this background. According to the standard random matrix terminology, the inverse temperature of this Coulomb gas equals 2, which means that the square of the absolute value of the Vandermonde determinant $\prod_{1 \le i<j \le T} |z_i-z_j|^2$ enters the probability measure.

As is well known from the random-matrix or the Coulomb-gas theory \footnote{Note, that we propose here an informal sketch of argument while the rigorous treatments can be found e.g. in \cite{AHM1}, \cite{AHM2}.}, in the $T\to \infty, \hbar T=2\pi t $ limit, expectation $\rho$ of the normalized density of eigenvalues (or Coulomb charges)
$$
\rho(z, \bar z)=<\hat \rho(z,\bar z)> =\int \hat \rho(z,\bar z) d{\cal P}, \quad \hat \rho(z,\bar z) = 2\pi\hbar \sum_{i=1}^T \delta(x-x_i)\delta(y-y_i)
$$
is a minimizer of the energy functional
$$
E[\rho]=\frac{-1}{4\pi^2}\int  \rho(z,\bar z)\rho(z', \bar z')\log|z-z'|dx'dy'dxdy+\frac{1}{2\pi}\int \rho(z,\bar z)\left(z\bar z+V(z,\bar z)\right)dxdy .
$$
The latter is obtained from the energy function (\ref{GEnergy}) by replacing $\hat \rho$ with its expectation $\rho$. The minimizer is restricted by the condition
\begin{equation}
\int \rho(z,\bar z)dxdy=\hbar T=2\pi t ,
\label{time}
\end{equation}
which fixes the total particle number (or, equivalently, the matrix size).

From the extremum condition $\frac{\delta}{\delta\rho}(E[\rho]-\lambda\int\rho dxdy)=0$, where $\lambda$ is a Lagrange multiplier, it follows that the electrostatic potential of the minimizer
$$
P(z,\bar z,t)=\frac{-1}{\pi}\int \rho(z',\bar z')\log|z-z'|dx'dy'+z\bar z+V(z,\bar z)
$$
must be constant on its support $\Omega: \rho(z, \bar z)=0, z\not\in\Omega$. Then, from the Poisson equation $\Delta P(z,\bar z)=\rho(z, \bar z)-1$, we obtain that $\rho=1$ in $\Omega$ and vanishes elsewhere.

Therefore, in the double scaling limit, eigenvalues form a constant density ``droplet", and the minimization of $E[\rho]$ reduces to determination of equipotential boundary $\partial \Omega$ of this droplet. This is exactly electrostatic interpretation of the Hele-Shaw problem given in equations (\ref{contour}), (\ref{VSigma}), while the Coulomb charge (or matrix eigenvalue) support is nothing but the solution $\Omega(t)$ of the Hele-Shaw problem. Thus, in the random matrix approach to the monopole-driven Hele-Shaw problem (\ref{velocity}, \ref{LG}, \ref{boundary}), the size of matrices plays the role of ``physical" time (see eq. (\ref{time})), while the $t$-independent external harmonic potential $V$ sets the initial conditions \footnote{We recall that $V$ is determined by $t_k,\bar t_k$ (see eq. (\ref{GEnergy})) which are in turn equal to the non-zero harmonic moments (\ref{HI}) that are constants of motion in this problem.}.

The above approach to statistical mechanics of Coulomb gases from the point of view of integrable hierarchies is not unique. Another, direct approach based on infinite soliton solutions, has been introduced in works \cite{LS}, \cite{LS_1D}, \cite{LS_2D} and later developed in \cite{Z1}.

First, we illustrate this approach on a simpler model of the one-dimensional lattice gas related to the KdV hierarchy.

As has been already mentioned, solutions of integrable hierarchies can be expressed by means of their $\tau$-functions \cite{MJD}. For instance, solution ${\cal V}={\cal V}(x, t_3, t_5, t_7, \dots)$ of the KdV hierarchy, with the Korteveg-de-Vriez equation
$$
\frac{\partial \cal V}{\partial t_3}+\partial_x^3{\cal V}-6{\cal V}\partial_x{\cal V}=0
$$
being the first one in the hierarchy, can be expressed through its $\tau$-function as
$$
{\cal V}(x,t_3,t_5,\dots)=-2\partial_x^2 \log \tau(x,t_3,t_5,\dots) .
$$
For a special class of solutions describing propagation of $N$ solitary waves, i.e. $N$-soliton solutions, the $\tau$ functions can be written down in the Hirota's form (see e.g. \cite{MJD})
\begin{equation}
\tau_N=\sum_{\sigma_1, \sigma_2, \dots \sigma_N}\exp\left(-\sum_{1\le l<l'\le N}G_{l,l'}\sigma_l\sigma_{l'}-\sum_{l=1}^N \sigma_l \theta_l\right)
\label{N-Soliton}
\end{equation}
where $\sigma_l$ take values 0 or 1. Soliton phases $\theta_l$ and ``phase shifts" $G_{l,l'}$ depend on soliton momentums $k_1, k_2, \dots k_N$ as
\begin{equation}
G_{l,l'}=-\log\frac{(k_l-k_{l'})^2}{(k_l+k_{l'})^2}, \quad \theta_l=-\phi_l-k_lx+k_l^3t_3+k_l^5t_5+k_l^7t_7+\dots
\label{Potentials}
\end{equation}
For instance, the one-soliton solution
$$
\tau_1=1+e^{-\theta_1}=1+e^{\phi_1+k_1x-k_1^3t_3-\dots}
$$
corresponds to uniform motion of a solitary wave with velocity $k_1^2$.

The two-soliton solution is a sum of 4 terms
$$
\tau_2=1+e^{-\theta_1}+e^{-\theta_2}+e^{-G_{12}-\theta_1-\theta_2}
$$
describing propagation of a pair of solitary waves, with asymptotic velocities $k_1^2$ and $k_2^2$, as well as a relative shift in their position acquired during their interaction.

Since the $N$-soliton $\tau$-function (\ref{N-Soliton}) contains $2^N$ terms depending on $N(N-1)/2$ pairwise interaction parameters $G_{ll'}$, it can be considered as a Grand partition function of particles on a lattice.
Indeed, it corresponds to a model on the lattice consisting of $N$ cites, each of which is either empty or occupied by at most one particle. Particles are interacting through two-particle potential $G_{l,l'}$ and with external potential $\theta_l$. In this interpretation, $\sigma_l$ stands for the filling factor of the $l$th site (which equals $1$ if there is a particle on that site and zero otherwise). The energy function of the lattice gas is
$$
E(\sigma_1, \sigma_2, \dots \sigma_N)=\sum_{1\le l<l'\le N}G_{ll'}\sigma_l\sigma_{l'}+\sum_{l=1}^N \sigma_l \theta_l, \quad \sigma_l \in \{0,1\},
$$
while the total number of particles equals
$$
n(\sigma_1, \sigma_2, \dots \sigma_N)=\sum_{l=1}^N \sigma_l.
$$
Note that the $N$-soliton $\tau$-function (\ref{N-Soliton}) can be also written down in a determinant form \cite{MJD} which allows one to find analytic solutions to physically interesting models.

Let us now impose condition of translational invariance on the two-particle interaction
$$
G_{l,l'}=U(l-l'),
$$
which, according to (\ref{Potentials}) is possible only if soliton momentums $k_l$ form a geometric series
$$
k_l=Ce^{2\hbar l}, \quad U(l)=-2\log\tanh|\hbar l|.
$$
The external one-particle interaction potential $\theta_l$ (which includes chemical potential of the gas) can be arbitrarily chosen by an appropriate setting of initial phases $\phi_l$, hierarchy ``times" $x, t_1, t_3, \dots $ and parameter $C$.

The limit $\hbar\to 0$ describes $\log$-gas on the uniform lattice, while in the double-scaling $N\to\infty$, $\hbar\to 0$ limit one gets $\log$-gas on the line with the following Grand partition function
$$
\tau=\sum_{n=0}^{N} \frac{1}{n!} {\cal Z}_n, \quad  {\cal Z}_n=\int \prod_{i=1}^n e^{-E_n(l_1, l_2, \dots l_n)} dl_i
$$
where
$$
E_n(l_1, l_2, \dots l_n)=-2\sum_{1\le i<j\le n}\log|l_i-l_j|-\sum_{i=1}^n \theta(l_i).
$$
In terminology of the random matrix theory, the temperature of this gas equals 2.

Grand partition functions of two-dimensional Coulomb systems could be obtained as multi-soliton solutions of hierarchies ``in 1+2 dimensions", such e.g. as Kadomtsev-Petviashvili (KP) or B-type KP hierarchy etc.

For example, in the case of $N$-soliton solution of the KP hierarchy, phase shifts and phases are determined by the soliton momentums $z_l, \bar z_l, l=1..N$ and a set of the hierarchy "times" $t_1, t_2, t_3, \dots$
$$
G_{l,l'}=-\log\frac{(z_l-z_{l'})(\bar z_l-\bar z_{l'})}{(z_l-\bar z_{l'})(\bar z_l-z_{l'})}, \quad \theta_l=\phi_l+(z_l+\bar z_l)t_1+(z_l^2-\bar z_l^2)t_2+(z_l^3+\bar z_l^3)t_3+\dots, \quad l=1..N.
$$
If we choose $z_l$ to be coordinates of a set of $N$ points in the plane (i.e. $z_l=x_l+iy_l, \bar z_l=x_l-iy_l$), then the $\tau$-function (\ref{N-Soliton}) becomes grand partition function of a two-dimensional lattice Coulomb gas in the upper half-plane with the following energy function of $n$-particle system
$$
E_n(\zeta_1, \zeta_2, \dots \zeta_n)=-2\sum_{1\le l<l'\le n} U(\zeta_l, \bar \zeta_l ; \zeta_{l'}, \bar \zeta_{l'}) +\sum_{l=1}^n \left( \phi(\zeta_l,\bar \zeta_l)+V(\zeta_l, \bar \zeta_{l'})\right), \quad \zeta_l\in\{z_1, \dots, z_N\}.
$$
The two-particle potential
$$
U(z, \bar z; z', \bar z')=-2\log|z-z'|+2\log|\bar z - z'|
$$
is the Coulomb potential in the upper half-plane $\mathbb{H}: {\rm Im}(z)>0$ , i.e. it equals to the fundamental solution of the Laplace equation $\Delta U(z, \bar z; z', \bar z')=2\pi \delta(x-x')\delta(y-y')$ that vanishes on the real axis $y={\rm Im}(z)=0$.

An external, one-particle potential consists of the harmonic part $V(z, \bar z)=\sum_{k=1}^\infty (z^k-\bar z^k)t_k$, that also vanishes at $y=0$, and the confining potential $\phi(z, \bar z)$, a such that $\phi(z_l, \bar z_l)=\phi_l$.

The model describes a logarithmic Coulomb gas at the inverse temperature 2
in the presence of an ideal conductor at $y$=0, so that each particle
interacts not only with other particles of the gas but also with their ``mirror images"
of the opposite charge.

In the limit of small lattice spacing $n$-particle partition function takes the form
$$
{\cal Z}_n=\int e^{-E_n(z_1, z_2, \dots z_n)} \prod_{l=1}^n \omega(z_l,\bar z_l)dx_ldy_l,
$$
where $\omega(z, \bar z)$ is a density of lattice cites (i.e. a number of lattice sites per unit area). Including this density in one-particle potential $\phi\to \phi+\log(\omega)$, one may set $\omega=1$.

\begin{figure}
\centering
\includegraphics[width=105mm]{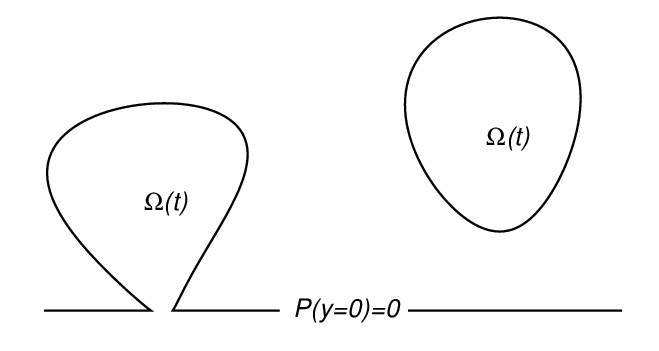}
\caption{A ``Fat slit" domain (left) and an isolated domain (right) growing in the upper half-plane $\mathbb H$. A viscous fluid in $\mathbb H\backslash\Omega(t)$ is pushed through constant pressure outlet located along the real axis $y=0$.} \label{KP}
\end{figure}

By analogy with the normal random matrix case (\ref{GEnergy}), a special limit of infinite-soliton solutions of the KP hierarchy corresponds to the Laplacian growth in the upper half-plane. This is a restricted Hele-Shaw problem with the constant pressure outlet along the real axis $P(x,0,t)=0$. Depending on choice of one-particle potentials, it can describe growth of isolated domain(s) or a ``fat slit" \cite{Z1,Z2} that runs through the outlet (see Figure \ref{KP}) or a combination of both.

Other types of integrable hierarchies are connected with various Hele-Shaw flows restricted by linear or circular impenetrable walls and/or constant pressure outlets \cite{LS_2D}, \cite{Z1}.
An unrestricted Hele-Shaw problem and the related grand partition function of normal matrix model are recovered in the limit when isolated domains are placed far away from walls and outlets.

\end{section}

\begin{section}{Elliptic Growth, Quadrature Domains and Algebraically Integrable Systems}
\label{Elliptic}

Above, we considered analytic approaches to the interface dynamics based on the theory of infinite-dimensional classical integrable systems. It turns out that not only hierarchies of integrable PDEs, but also finite dimensional quantum integrable systems emerge in analytic study of interface dynamics. These include algebraically integrable systems connected with integrable elliptic growth, being a generalization of the Hele-Shaw problem that describes diffusive growth processes in non-homogeneous media.

Algebraically integrable systems are non-trivial generalizations of the Laplace operators that have been mainly studied in connection with unsolved Hadamard's problem for hyperbolic PDEs (for a review see e.g. \cite{BV, BL}). Theory of algebraically integrable systems was applied to elliptic growth and Hadamard's problem in works \cite{L, LY1, HLO, BdMLY}.

In the elliptic growth problem, boundary dynamics obeys the following law
$$
v_n=\kappa(x,y)\partial_n P(x,y,t),
$$
which is a generalization of (\ref{velocity}), while the pressure satisfies the following variable-coefficient elliptic equation
\begin{equation}
\nabla \kappa\eta \nabla P(x,y,t)=\sum_{i=1}^{k+1} q_i(t)\delta(x-x_i)\delta(y-y_i), \quad \kappa>0, \eta>0.
\label{EG}
\end{equation}
In the theory of porous medium flows, $\kappa(x,y)$ is referred as permeability or conductivity, and $\eta=\eta(x,y)$ is medium porosity.

The above two equations, together with zero pressure condition (\ref{boundary}) on the moving boundary, constitute a generalization of the Laplacian growth (Hele-Shaw) problem, which is recovered as particular case with $\eta={\rm const}, \nu={\rm const}$ of elliptic growth (for more detailed introduction to elliptic growth see \cite{KMP, Lu}).

The term ``elliptic growth" first appeared in \cite{KMP}. Obviously, different aspects of the elliptic growth have been studied before under the names ``variable coefficient Hele-Shaw problem", ``free-boundary inhomogeneous porous medium flows", ``Non-Laplacian growth" etc. In particular, the first example of integrable elliptic growth appeared in work \cite{E}.

When all sources/sinks are placed close to each other, (\ref{EG}) becomes
$$
\nabla \kappa\eta \nabla P(x,y,t)=\hat q[\delta(x-x_1)\delta(y-y_1)],
$$
where $\hat q$ is the differential operator of order $k$
$$
\hat q=q_0(t)+\sum_{j=1}^k (-1)^j\left(
q_j(t)\frac{\partial^j}{\partial z^j}+\bar
q_j(t)\frac{\partial^j}{\partial \bar z^j}\right).
$$
In practice one wishes to control the boundary dynamics using a finite number of closely placed fixed sources/sinks by varying their intensity in time. In the case of the interior problem, one may control boundary dynamics by setting up a time dependence of the coefficients of the analytic (in $\Omega$) polynomial conformal mapping
\begin{equation}
f(w,t)=z_1+r(t)w+\sum_{l=2}^k u_l(t)w^l, \quad \{w: |w|=1\} \to \partial\Omega(t).
\label{Polynomial}
\end{equation}
Then, it is necessary to find such $q_i(t)$ that the solution of corresponding elliptic growth problem would be as close as possible to $\Omega(t)$ of (\ref{Polynomial}).

The Laplacian growth possesses a remarkable property that dynamics corresponding to polynomial mapping of a fixed degree (\ref{Polynomial}) can be always sustained exactly by a finite number of sources. So, it is natural to ask how to describe the elliptic growth problems having similar property? In more detail: For which medium (i.e. for which $\kappa$ and $\eta$) the elliptic growth reproduces any \footnote{This inverse problem is different from the one posed in \cite{KMP} where $\kappa$ and $\eta$ are fitted for a given ``movie", while sources are unchanged. In our problem $\kappa$ and $\eta$ (i.e. characteristics of medium) are ``movie-independent".} polynomial dynamics ``movie" of the Laplacian growth at the expense of adding a finite number of the multipole sources at $z=z_1$? Complete classification of such media is a difficult unsolved problem, but it turns out that there is a wide class of processes related to the algebraically integrable systems that satisfy such a requirement.

In the Laplacian growth case, one could derive the polynomial dynamics property either through the Polubarinova-Kochina equation (\ref{LG}) or by using conservation laws in form of quadrature identities. Due to the fact that a generic elliptic equation is not invariant under arbitrary conformal transformations, we are restricted to application of the latter approach. In more details, the following quantities
$$
I[\phi]=\int_{\Omega(t)}\eta\phi dxdy,
$$
where $\phi=\phi(x,y)$ is any regular in $\Omega$ time-independent homogenous solution of the elliptic PDE
\begin{equation}
\nabla \kappa \eta \nabla \phi =0,
\label{elliptic}
\end{equation}
have simple time dynamics:
$$
\frac{dI[\phi]}{dt}=q_0(t)\phi(x_1,y_1)+\sum_{j=1}^k\left(
q_j(t)\frac{\partial^j\phi(x,y)}{\partial z^j}+\bar
q_j(t)\frac{\partial^j\phi(x,y)}{\partial \bar z^j}\right)_{z=z_1}.
$$
The time derivative of $I[\phi]$ depends only on a value of $\phi$ and a finite number of its derivatives evaluated at the fixed point $z=z_1=x_1+iy_1$. An infinite number of invariants of motion can be obtained by choosing $\phi(x,y)$ that make the rhs of the above equation vanish. For example, in the Hele-Shaw problem (\ref{velocity}), (\ref{LG}), (\ref{boundary}), these are harmonic moments (\ref{HI}).

In the case when liquid is injected in an initially empty medium, $I|_{t=0}=0$ and the following quadrature identity holds
\begin{equation}
\int_{\Omega(t)}\eta\phi dxdy=\hat Q[\phi](x_1,y_1), \quad \hat Q=Q_0+\sum_{j=1}^k \left(
Q_j\frac{\partial^j}{\partial z^j}+\bar
Q_j\frac{\partial^j}{\partial \bar z^j}\right)
\label{Quadrature}
\end{equation}
for any solution of (\ref{elliptic}). The coefficients of differential operator $\hat Q=\hat Q(t)$ do not depend on a history of sources $\hat q(t)$ and are equal total fluxes injected by time $t$
$$
Q_j(t)=\int_0^t q_j(t')dt', \quad j=0..k .
$$
The quadrature identity (\ref{Quadrature}) is a generalization of the mean-value theorem for harmonic functions. Special domains for which quadrature identities hold are called quadrature domains (for more information on the quadrature domains see \cite{GS} and references therein and \cite{Cr} for survey of connections to problems of fluid dynamics).

The idea of derivation of the polynomial dynamics property is to prove that identity (\ref{Quadrature}) holds in domains given by (\ref{Polynomial}). This derivation may be reduced to a residue calculus, if the kernel elements of the elliptic operator
\begin{equation}
L=\nabla\xi^{-2}\nabla
\label{L}
\end{equation}
can be expressed through action of a differential operator $T$ with rational in $z, \bar z$ coefficients on harmonic functions, i.e.
\begin{equation}
L\phi=0, \quad \phi=T[f(z)+g(\bar z)],
\label{TL}
\end{equation}
where $g(\bar z), f(z)$ are arbitrary (anti)analytic in $\Omega$ functions.

A subclass of algebraically integrable systems provides families of such operators. In more details, a $d$-dimensional Schrodinger operator
\begin{equation}
H=\xi(\zeta) L \xi(\zeta) = \xi\nabla\xi^{-2}\nabla\xi = \Delta + {\cal V}(\zeta), \quad {\cal V} =-\xi\Delta \left[ \xi^{-1}\right]=\frac{\Delta\xi}{\xi}-2\left(\frac{\nabla\xi}{\xi}\right)^2,
\label{Schrodinger}
\end{equation}
where $\zeta:=(\zeta_1, \dots \zeta_d)$, is algebraically integrable if it is an element of an over-complete ring of commuting linear differential operators, i.e. a ring whose minimal set of generators consists of more than $d$ operators (commuting integrals).

\begin{figure}
\centering
\includegraphics[width=77mm]{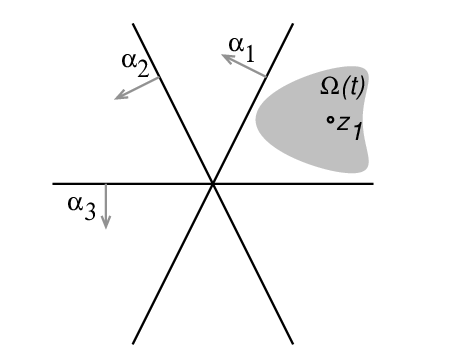}
\caption{Example of a linear singular locus of multi-dimensional Baker-Akhieser function in two dimensions with three hyperplanes (lines).} \label{Loci}
\end{figure}

The notion of algebraic integrability is linked with the concept of multi-dimensional Baker-Akhieser function \cite{CFV}, \cite{CSV}, which is determined completely by its singular locus: a set of $n$ hyperplanes of dimension $d-1$ with normals $\alpha_1, \dots \alpha_n $ and multiplicities $m_1, \dots m_n \in \mathbb N$ (see Figure \ref{Loci}). The hyperplanes may be situated in linear or affine arrangement. This is a function that depends on spatial coordinates $\zeta:=(\zeta_1,\dots \zeta_d)$ and ``momentum" spectral coordinates $k=(k_1, \dots, k_n)$ as
$$
\psi(\zeta,k)=\frac{R(\zeta,k)}{A(k)}e^{(k\cdot\zeta)},
$$
where $A(k)=(k\cdot\alpha_1)(k\cdot\alpha_2)\dots(k\cdot \alpha_n)$ and $R(\zeta,k)$ is a polynomial in $k$ with the highest term being $A(k)$. The latter polynomial is fixed by the following conditions
$$
\partial_\alpha\psi(\zeta,k)(k\cdot\alpha)^{m_\alpha}=\partial_\alpha^3\psi(\zeta,k)(k\cdot\alpha)^{m_\alpha}=\dots=\partial_\alpha^{2m_\alpha-1}\psi(\zeta,k)(k\cdot\alpha)^{m_\alpha}=0,
$$
where $\partial_\alpha=(\alpha\cdot\partial/\partial k)$, on the hyperplane $(k\cdot\alpha)=0$ for all $\alpha\in\{\alpha_1,\dots\alpha_n\}$.

If the Baker-Akhieser function exists, then it is unique and satisfies the algebraically integrable Schrodinger Equation \cite{CSV}, \cite{CFV}
$$
H[\psi](\zeta,k)=k^2\psi(\zeta,k),
$$
with the Schrodinger operator (\ref{Schrodinger}) having the following potential
$$
{\cal V}(\zeta)=-\sum_{\alpha}\frac{m_\alpha(m_\alpha+1)(\alpha\cdot\alpha)}{(\alpha\cdot\zeta)^2}
$$
which is singular on the hyperplanes $(\alpha\cdot\zeta)=0$ (see Figure \ref{Loci}).

The Baker-Akhieser function exists only for very special arrangements of singular hyperplanes. The problem of complete classification of such singular loci is far from being solved.

The rational Baker-Akhieser function automatically implies that there is a differential operator ${\cal T}$ with rational coefficients (an ``intertwining" operator) that connects $H$ with Laplacian by the following relation
$$
H{\cal T}={\cal T}\Delta, \quad {\cal T}=R(\zeta, \nabla).
$$
Therefore, algebraically integrable systems equip us with operator $T$ in equation (\ref{TL})
$$
T=\xi(\zeta){\cal T}
$$
needed for the proof of quadrature identities in polynomial domains.

It is not clear if algebraic integrability alone leads to existence of polynomial dynamics in the related elliptic growth problem. The existence has been shown for classes of systems related to finite-reflection (Coxeter) groups, rational solutions of the KdV hierarchy \cite{BdMLY, L, LY1} as well as for a subclass of a class related to $2\pi$-periodic $n$-soliton solutions of the KdV hierarchy \cite{BL}.

In the case of the Coxeter configurations on the plane
\begin{equation}
H=\Delta+2\Delta[\log\tau],
\label{HTau}
\end{equation}
where
$$
\tau=(z^l-\bar z^l)^{m_1(m_1+1)/2}(z^l+\bar z^l)^{m_2(m_2+1)/2}, \quad l, m_1, m_2 \in \mathbb N
$$
corresponds to the elliptic growth problem with the permeability and porosity
$$
\kappa=\xi^{-2}=(z^l-\bar z^l)^{-2m_1}(z^l+\bar z^l)^{-2m_2}, \quad \eta=1
$$
that are invariant wrt reflections in $l$ or 2$l$ singular hyperplanes.

For the class of integrable elliptic growth related to rational solutions of the KdV hierarchy, $\tau$ in (\ref{HTau}) is a special limit of the $\tau$-function of the hierarchy that can be expressed in terms of Adler-Moser (Burchnall-Chaundy) polynomials $p_l(x)$ (see e.g; \cite{AM, BC}). The $l$th Adler-Moser polynomial depends on $l-1$ non-trivial free parameters $t_3, t_5, \dots t_{2l-1}$ and obeys the bilinear recurrence relation:
$$
\tau=p_l(x, t_3, t_5, \dots t_{2l-1}), \quad p_0=1, p_1=x,
$$
$$
p_{l+1}^{\prime\prime}(x)p_l(x)-2p_{l+1}^\prime(x)p_l^\prime(x)+p_{l+1}(x)p_l(x)^{\prime\prime}=0.
$$
Note that in difference from the Coxeter case, a generic system of the present class has an affine singular locus and
describes flows in stratified media, i.e. the media with permeability and porosity changing only in one (e.g. $x$) direction
$$
\kappa\eta=\xi^{-2}, \quad \xi(x,y)=\frac{p_n(x)}{p_{n-1}(x)}, \quad \eta(x,y)=p_{n-1}(x)S(x),
$$
where $S$ is an arbitrary polynomial of $x$.

As has been mentioned above, there are also several ``non-standard" examples related to soliton solutions of the KdV hierarchy. For instance, the following ``$\tau$-function" in (\ref{HTau})
$$
\tau=x^m((2m+1)y^2-x^2), \quad m\in \mathbb N
$$
corresponds to the integrable elliptic growth with
$$
\kappa=\xi^{-2}=x^{-2m}((2m+1)y^2-x^2)^{-2}, \quad \eta=1.
$$
This is the first non-trivial example (found in \cite{CFV1}) of the linear singular loci that are not reflection invariant (except for the case $m=1$, which corresponds to the Coxeter group $A_2$). It can be considered as one of the particular cases within the algebraically integrable class introduced in works on (still unsolved) Hadamard's problem (see e.g. \cite{BL}).

Concluding this section we would like to mention results on integrable elliptic growth in dimensions higher than two. It is of interest either for description of three-dimensional porous medium flows, or from the point of view of generalization of harmonic analysis in arbitrary dimensions.

Similarly to the planar case, the elliptic growth in higher dimensions also possesses an infinite number of conservation laws and quadrature identities hold in an arbitrary dimension. However, the conformal mapping technique is unavailable for $d>2$ and alternative approaches are needed for proof of such identities.

It is worth to note that the problem of higher-dimensional quadrature domains is challenging and interesting already in the more basic context of Laplacian Growth. Here one can mention the question by Shapiro \cite{Sh} on algebraicity of quadrature domains in more than two dimentions as well as some non-trivial examples of Laplacian growth in four dimensions \cite {EL}, \cite{K}, \cite{Lu}. A physically important problem that remains open is to construct nontrivial quadrature domains in three dimensions.

In the case of the elliptic growth related to quantum algebraically integrable systems one can use a direct approach  to the problem: One finds such $q_i(t)$ and $P(\zeta,t)$ that solve elliptic growth problem for a chosen domain. The progress here is achieved only for simplest examples, i.e. for spherical fronts \cite{BdMLY}. Existence of such a solution implies quadrature identity for spheres (balls), which is a generalization of the mean-value theorem for harmonic functions to elliptic operators with variable coefficients for an arbitrary $d$.

Here, the solution procedure involves application of methods developed for construction of fundamental solutions of algebraically integrable operators \cite{BM}. By equation (\ref{EG}) ($\hat q$ now denotes a differential operator corresponding to multipole expansion of the $d$-dimensional source)
$$
P(\zeta,t)=\hat q [G](\zeta, \zeta')+\Psi(\zeta,\zeta'),
$$
where $G$ is a fundamental solution of the elliptic operator (\ref{L}) and $\Psi(\zeta, \zeta')$ is its particular homogenous solution:
$$
L[G](\zeta, \zeta')=\delta^d(\zeta-\zeta'), \quad L[\Psi](\zeta,\zeta')=0.
$$
The fundamental solution $G$ can be built by means of the Hadamard expansion. This is an expansion in terms of fundamental solutions $\Delta^{-i}G_0$ of powers of Laplacian $\Delta^{i+1}$. The algebraic integrability leads to truncation of this expansion and, after an appropriate choice of $\Psi$, to solution of the problem.

\end{section}

\begin{section}{Laplacian Growth as Iterative Stochastic Conformal Mappings: Exact Results}
\label{LLE}

One of the main direct goals of the theory of stochastic Laplacian growth is determination of multi-fractal characteristics of the interface. So far, the only non-trivial examples where full analytic description of such characteristics was found are special cases of Stochastic Loewner Evolution.

The Levy-Loewner Evolution (LLE) and its special case Schramm-Loewner evolution (SLE) are stochastic processes that are close ``cousins" of the Laplacian Growth/DLA: In all the above models the interface dynamics is governed by harmonic scalar field and can be represented by iterative conformal mappings. However, in difference from DLA, the SLE/LLE are conformally invariant and, in a sense, are linear stochastic processes. The latter properties help to obtain exact multi-fractal characteristics of SLE and facilitate analysis in the case of LLE.

We first introduce the radial Levy-Loewner evolution
(a good introduction to the chordal LLE can be found in
\cite{ORGK}, \cite{ROKG}, for a quick introduction to Levy processes see e.g. \cite{A} and references therein).
The iterative conformal map representation of DLA \cite{HL} and its relationship with LLE will be brifely discussed after.

\begin{figure}
\centering
\includegraphics[width=115mm]{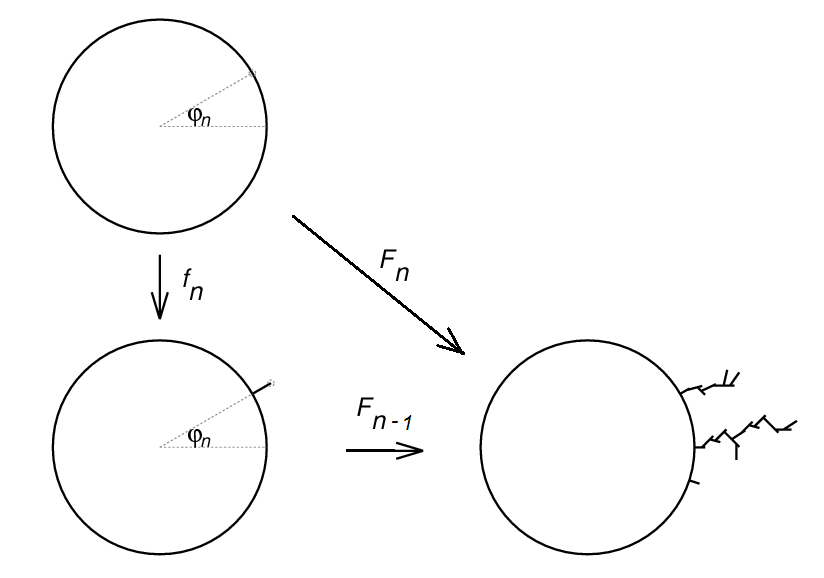}
 \caption{\small The elementary mapping $z=f_n(w, \delta t_n)$ (left) and the compound iterative mapping $z=F_n(w)$ (right) from exterior of unit circle in $w$-plane to exterior of a simply connected domain in $z$-plane.}
 \label{HL_LLL}
 \end{figure}

Let us consider iterative conformal mappings $z=F_n(w)$ from an
exterior of the unit disc in the $w$-plane to an exterior of a
bounded, simply connected domain in the $z$-plane: The $n$th mapping
is a composition of $n$ elementary ``spike" mappings $z=f_j(w,
\delta t_j)$, $j=1..n$
\begin{equation}
F_{n}(w)=F_{n-1}(f_n(w, \delta t_n)), \quad F_0(w)=w,
\label{Fn}
\end{equation}
where
\begin{equation}
f_n(w, t)=e^{i\varphi_n}h\left(e^{-i\varphi_n} w, t\right), \quad
h(w, t)=e^{ t}(w+1)\frac{w+1+\sqrt{(w+1)^2-4e^{- t}w}}{2w}-1 \label{atom}
\end{equation}
The elementary mapping $z=f_n(w, \delta t_n)$ attaches a radial
``spike" of the length $\sqrt{\delta t_n}\left(1+ O(\delta
t_n)\right)$  located the angle $\varphi_n$ to unit disc:
Here, the point $w=e^{i\varphi_n}$ on the unit circle in the
$w$-plane is mapped to the tip of the spike in the $z$-plane (see
Figure \ref{HL_LLL}).

The mapping $z=h(w, t)$ that attaches a spike to the disc at
$\varphi=0$ satisfies the simplest Loewner equation $ \frac{\partial
h(w, t)}{\partial t}=w\frac{\partial h(w, t)}{\partial
w}\frac{w+1}{w-1} $ and as a consequence
$$
\frac{\partial f_n(w, t)}{\partial t}=w\frac{\partial f_n(w,
t)}{\partial w}\frac{w+e^{i\varphi_n}}{w-e^{i \varphi_n}}, \quad f_n(w,0)=w.
$$
This equation is invariant wrt any conformal transformation $f_n(w,
t)\to F\left(f_n(w, t)\right)$ and therefore an iterative compound
mapping $F_n$ in (\ref{Fn}) can be represented as a solution of the
Loewner equation at time $t=\sum_{j=1}^n \delta t_j$:
$$
F_n(w)=F\left(w, \sum_{j=1}^n \delta t_j\right), \quad F(w,0)=w,
$$
\begin{equation}
\frac{\partial F(w, t)}{\partial t}=w\frac{\partial F(w,
t)}{\partial w}\frac{w+e^{i L(t)}}{w-e^{i L(t)}} .
\label{LE}
\end{equation}
Here $L(t)$ is the piecewise constant function
$$
L(t)=\varphi_j, \quad \sum_{k=1}^{j-1} \delta t_k < t < \sum_{k=1}^j
\delta t_k
$$
One can consider continuous-time Loewner evolutions as limits of
the above iterative processes when $n\to\infty$ and $\delta t_n\to
0$.

We are interested in the case when $L(t)$ is a stochastic process
without a drift. Without loss of generality we set
\begin{equation}
L(0)=0, \quad \langle L(t) \rangle = 0,
\label{Drift=0}
\end{equation}
where $\langle \rangle$ denote expectation (ensemble average).

When a stochastic process $L(t)$ is continuous in time, the conformal
mapping $z=F(w,t)$ describes growth of a random continuous curve
$\Gamma=\Gamma(t)$ starting from a point on a unit circle $|z|=1$ at
$t=0$. On the other hand, when $L(t)$ is discontinuous in time, the
growth branches in the $z$-plane.

If one requires that the Loewner evolution (\ref{LE}) is a
conformally invariant Markovian process, in a sense that the time
evolution is consistent with composition of conformal maps \footnote{which implies that the probability distribution of $z=F^{-1}\left(F(w,t+\tau),t\right)$ coincides with that of $z=F(w,\tau)$, where $w=F^{-1}(z,t)$ is an inverse of $z=F(w,t)$},
then the necessary condition for such an evolution is that $L(t)$ must have independent stationary
increments, i.e. $L(t)$ is a Markovian process with the probability distribution of
$L(t+\tau)-L(t)$ depending only on $\tau$, i.e. a Levy process.

The only continuous (modulo uniform drift) process of Levy type is
the Brownian motion
\begin{equation}
L(t)=B(t)
\label{B}
\end{equation}
The Brownian motion is characterized by a single parameter -
``temperature" $\kappa$ :
\begin{equation}
\langle (B(t+\tau)-B(t))^2 \rangle = \kappa |\tau| . \label{Bt}
\end{equation}

The stochastic Loewner evolution driven by Brownian motion is called
Schramm-Loewner Evolution (SLE, or SLE$_\kappa$). Since it describes
non-branching planar stochastic curves with a conformally-invariant
probability distribution, SLE is a useful tool for description of
boundaries of critical clusters in two-dimensional equilibrium
statistical mechanics. In this picture, different $\kappa$
correspond to different classes of models of statistical mechanics
(a good introduction to SLE for physicists can be found e.g. in \cite{C}, \cite{G} as
well as mathematical reviews are given e.g. in \cite{GL},
\cite{GL2}).

Now let us briefly discuss similarities and differences between DLA and LLE: Similarly to LLE, the DLA can be also represented by iterative conformal mappings shown on Figure \ref{HL_LLL}. Moreover, these mapping are driven by very simple process: the probability for angle $\varphi_n$ is uniformly distributed along the unit circle and is independent of $\varphi_{n-1}$. However, in difference from LLE, where conformal mapping satisfies the linear PDE (\ref{LE}), evolution in the case of DLA is non-linear.

In more detail, addition of a spike during an iteration corresponds to attachment of a particle to a cluster in DLA (see section \ref{LG_DLA}). Since the area of the particle is constant the parameter $\delta t_n$ of the elementary mapping in (\ref{Fn}) now depends on $F_{n-1}$ as follows
$$
\delta t_n = \delta a \left|\frac{\partial F_{n-1}}{\partial w}\right|^{-\alpha}_{w=\exp i\varphi_n}
$$
where $\delta a$ is constant and $\alpha=2$ in the case of DLA. As a consequence, description of DLA does not simplify in this approach. The model with an arbitrary $\alpha$ is called the Hastings-Levitov model or HL$(\alpha)$. It simplifies in the $\alpha=0$ case, but due to triviality of probability distribution of $\varphi_n$ and $\delta t_n$ in the HL models, we get trivial multi-fractal spectrum when $\alpha=0$: Despite unlimited branching in HL$(0)$ process, the multi-fractal spectrum of the HL$(0)$ coincides with that of a regular smooth one-dimensional curve \cite{RZ}.

Although generic LLE is governed by a linear PDE, one can consider non-trivial Levy processes, such as e.g. Brownian motion. As mentioned above, one gets non-trivial exact results for the latter sub-class of LLE (i.e. for SLE).

One usually considers the ``bounded whole plane" LLE that is a properly scaled infinite-time limit of the radial LLE
\begin{equation}
\mathcal{F}(w,t)=\lim_{T\to\infty}e^{-T}F(w,T+t),
\label{wp+ext}
\end{equation}
which describes growth of stochastic curve out of the point of
origin in the plane (rather than out of the circle). Obviously, $\mathcal{F}(w,t)$ also satisfies the radial Loewner equation(\ref{LE}) with time running from $t=-\infty$ to $t=\infty$.

The ``unbounded" version of the whole-plane LLE is an
inversion $\mathcal{F}(w,t)\to 1/\mathcal{F}(1/w,t)$ of the bounded version. Obviously, this mapping from the unit disc in the $w$-plane to the complement of stochastic curve which grows from infinity towards the origin in the $z$-plane also satisfies the Loewner equation (\ref{LE}). It has been studied mainly due to its relationship with the problem of Bieberbach
coefficients of conformal mappings \cite{DNNZ}, \cite{LSLE},\cite{LYLLE}.

Let us briefly recall what are multi-fractal spectra and how different types of such spectra are related with each other (For further introduction to multi-fractal analysis see e.g.
\cite{BS}, \cite{BDZ}, \cite{HJKPS}, \cite{H}, \cite{HHD}, \cite{Wiki} and references therein). Suppose that some measure is defined on our interface/curve (this can be electric flux, probability, mass, magnetization etc). Then dividing the space into boxes of size $l$, we look for the distribution of scalings of the measure wrt change of $l$ inside the boxes containing the curve.

For instance, consider the harmonic measure of a planar curve at some point $z=\mathcal{F}(w)$, $w=e^{i\varphi}$, which is proportional to the probability of a random walker released from infinity to hit the curve at $z$. This probability is, in turn, inversely proportional to $|\mathcal{F}'(w)|$. For the curve covered with $N$ boxes of size $l$, we denote the probability of hitting the curve at the $j$-th box by $p_j$. The observable properties of the measure are expressed through the $\tau(q)$-spectrum
$$
\sum_{j=1}^N p_i^q \asymp l^{\tau(q)}.
\label{pq}
$$
It is easy to see that $-\tau(0)$ is the Hausdorff dimension of the curve. Another common definition is the generalized dimension $D(q)=\tau(q)/(q-1)$ with $D(0)$, $D(1)$ and $D(2)$ being the Hausdorff, information and correlation dimensions respectively.

To estimate the number of boxes for which probability scales as $l^{\alpha}$ with the change of $l$, one introduces the $f(\alpha)$-spectrum:
\begin{equation}
\sum_{i=j}^N p_j^q \asymp \int l^{-f(\alpha)}l^{q\alpha}d\alpha, \quad l\to 0
\label{falpha}
\end{equation}
As $l\to 0$, the integral in (\ref{falpha}) will be dominated by value of $\alpha$ which makes $q\alpha-f(\alpha)$ smallest and it follows that $\tau$-spectrum and $f$-spectrum are related by the Legendre transform
$$
\tau(q)=\inf_\alpha\left[q\alpha-f(\alpha)\right], \quad f(\alpha)=\sup_q\left[q\alpha-\tau(\alpha)\right] .
$$
To get the multi-fractal spectrum of LLE one has to find first the so-called ``$\beta$-spectrum": The integral means $\beta(q)$-spectrum of the
domain is defined through the $q$th moment of a derivative of conformal mapping at the unit
circle (i.e. at $|w|\to 1$) as follows
\begin{equation}
\beta(q)=\overline{\lim}_{\epsilon\to0+}\frac{\log
\int_0^{2\pi}\langle\left|\mathcal{F}'\left(e^{\epsilon+i\varphi}\right)\right|^q\rangle
d\varphi}{-\log \epsilon} .
\label{beta}
\end{equation}
The $\beta$-spectrum and $f$-spectrum are also related by the Legendre transform (see e.g. \cite{H,BS,BDZ,LYLLE}):
$$
\beta(q)=\sup_{\alpha}\left[q-1+(f(\alpha)-q)/\alpha\right], \quad
f(\alpha)=\inf_q\left[q+\alpha(\beta(q)+1-q)\right] .
$$
To find $\beta(q)$-spectrum (\ref{beta}) one needs to estimate moments of derivative: One can show that the value
$$
\rho=e^{\mp qt}\langle\left|\mathcal{F}'\left(e^{iL(t)}w,t\right)\right|^q\rangle,
$$
where $\pm$ signs correspond to bounded and unbounded whole-plane LLE, is time independent and is a function of $w,\bar w$ and $q$ only \cite{BS, BDZ, DNNZ, LSLE, LYSLE, LYLLE}. Moreover it satisfies the linear equation
\begin{equation}
L\rho=\pm q\rho \label{Lrho}
\end{equation}
where the linear operator $L$ equals
\begin{equation}
L=-\hat\eta+w\frac{w+1}{w-1}\partial_w+\bar w\frac{\bar w+1}{\bar
w-1}\partial_{\bar w}-\frac{q}{(w-1)^2}-\frac{q}{(\bar w-1)^2}+q
\label{Leta}
\end{equation}
In (\ref{Leta}), the linear operator $\hat\eta$ acts on functions of $w, \bar w$ as follows
\begin{equation}
\hat\eta[\rho](w, \bar w)=\lim_{t\to
0}\frac{1}{t}\int_0^{2\pi}\left(\rho(w, \bar
w)-\rho\left(e^{i\varphi}w, e^{-i\varphi}\bar
w\right)\right) P(\varphi, t) d\varphi,
\label{eta}
\end{equation}
where $P(\varphi,t)$ is the probability density that $L(t)=\varphi$ under
condition $L(0)=0$. In polar coordinates $(r,\phi)$ (such that $w=re^{i\phi}, \bar{w} =re^{-i\phi}$) and in terms of the symmetric (wrt reflection $\phi\to -\phi$) Levy measure \footnote{The Levy process can be viewed as a Brownian motion with jumps, where distribution of intensity and magnitude of jumps depends on the Levy measure $d\eta$.} $d\eta(\phi)\ge 0$, the action of operator $\hat\eta$ on a function $f(r,\phi)$ writes as
$$
\hat\eta [f](r,\phi)=-\frac{\kappa}{2}\partial_\phi^2f(r,\phi)+\int_{-\pi}^{\pi}\left(f(r,\phi)-f(r,\phi+\varphi)\right) d\eta(\varphi).
$$
One may say that the first (differential) term in the above equation corresponds to the continuous part of LLE, while the second (integral) term is related to the branching of LLE curves (For more details on equation (\ref{Leta}) see e.g. \cite{LYLLE, LYLLEFP}).

To find moments of derivative in the bounded case one has to look for a solution of (\ref{Lrho}) that is analytic in $w$ and $\bar w$ at infinity and $\rho\to 1$ as $w\to\infty$ (in the unbounded case similar conditions are imposed at $w=0$).  Such a solution is unique.

In the case of  Brownian motion (\ref{Bt}), $d\eta=0$ while $P(\varphi,t)$ in (\ref{eta}) is a
fundamental solution of the heat equation on the circle and
\begin{equation}
\hat\eta=-\frac{\kappa}{2}\partial_\phi^2=\frac{\kappa}{2}\left(w\partial_w-\bar w\partial_{\bar
w}\right)^2 ,
\label{eta_kappa}
\end{equation}
i.e. equation (\ref{Lrho}) becomes the second order linear PDE in $w$ and $\bar w$.
This allows to find exact (analytic) form of the multi-fractal spectrum of the Schramm-Loewner Evolution. For example, the $\beta$-spectrum of bounded whole-plane $SLE_\kappa$ is
$$
\beta(q)=\left\{\begin{array}{ll}
\kappa\frac{\gamma(q,\kappa)^2}{2}-2\gamma(q,\kappa)-1 , &  \quad q\le -1-\frac{3\kappa}{8} \\
\kappa\frac{\gamma(q,\kappa)^2}{2}, & \quad -1-\frac{3\kappa}{8} \le q \le \frac{3(\kappa+4)^2}{32\kappa} \\
q-\frac{(\kappa+4)^2}{16\kappa}, & \quad  q \ge \frac{3(\kappa+4)^2}{32\kappa}
\end{array}
\right.,
$$
where
$$
\gamma(q,\kappa)=\frac{\kappa+4-\sqrt{(\kappa+4)^2-8q\kappa}}{2\kappa}.
$$
It is worth to note that derivation of the multi-fractal spectrum based on the above simple method first appeared for chordal SLE in \cite{Has}. It was preceded by derivation based on the Conformal Field Theory and 2D quantum gravity in \cite{D}.

\end{section}

\begin{section}{Concluding Remarks}

In this paper we reviewed main advances in analytic study of the Laplacian growth (in wide sense). Among the exact results obtained in the theory of Laplacian Growth not reviewed here one might mention several other advances, such as an attempt of solution of the Saffman-Taylor finger selection problem with the help of exact ``$N$-finger" solutions of the Hele-Shaw problem \cite{M}, establishing of relationship between the Whitham hierarchy and the Hele-Shaw problems for multiply-connected domains \cite{KMWZ}, Schwartz function/potential approach to the Laplacian/Elliptic growth \cite{GTV, Lu, Sh}, Poisson growth \cite{KMP}, relationship of special types of the Laplacian Path Models with dynamics of classical Calogero-Sutherland-Moser systems \cite{CM,S}, multiple SLE and quantum Calogero-Sutherland systems \cite{C}, relationship between the LLE and theories of Fuchsian systems  and orthogonal polynomials \cite{LYLLEFP} as well as the problem  of the Bieberbach coefficients in LLE \cite{DNNZ, LYLLE} etc.

\end{section}


\begin{thebibliography}{101}

\bibitem {AM} M. Adler, J. Moser, {\it On a class of polynomials connected
with the Korteveg-de Vries equation}, Comm.Math.Phys. 61 (1978) 1-30


\bibitem{A} Applebaum D, {\it Levy Processes - From Probability to Finance and Quantum Groups}, Notices of the AMS,  51 (11): 1336–1347 (2004)

\bibitem{AHM1} Y. Ameur, H. Hedenmalm, and N. Makarov, {\it Random normal matrices and
Ward identities}, Ann. Probab., 43(3):1157-1201, 2015.

 \bibitem{AHM2} Y. Ameur, H. k. Hedenmalm, and N. Makarov, {\it Fluctuations of eigenvalues of
random normal matrices}, Duke Math. J., 159(1):31-81, 2011.

\bibitem{BS}
D. Beliaev, S. Smirnov, \emph{Harmonic measure and SLE}, Commun.
Math. Phys. 290, 577–595 (2009).

\bibitem{BDZ} D. Beliaev, B. Duplantier, and M. Zinsmeister, {\it Integral means spectrum of whole-plane SLE}, Comm. Math. Phys. Volume 353, Issue 1, pages 119–133, (2017).

\bibitem{BKLST} D. Bensimon, L. Kadanoff, S. Liang, B. Shraiman, C. Tang, {\it Viscous flows in two dimensions}, Rev. Mod. Phys. 58 (1986) 977.

\bibitem{BL} Y.Berest and I.Loutsenko,
{\it Huygens Principle in Minkowski Spaces and Soliton Solutions of the
Korteweg-de Vries Equation}, Comm.Math.Phys 190, (1997) 113-132

\bibitem{BM} Yu.Yu.Berest, Yu.A.Molchanov, {\it Fundamental Solutions for
Partial Differential Equations with Reflection Group Invariance},
J.Math.Phys., 36(8) (1995), 4324-4339

\bibitem{BV} Berest, A.P. Veselov, {\it Huygens' principle and integrability}, Russ. Math. Surv., 49(6), 5-77

\bibitem{BdMLY} A. Boutet de Monvel, I. Loutsenko, O. Yermolayeva, {\it New applications of quantum algebraically integrable systems in fluid dynamics}, Analysis and Mathematical Physics, v.3, Issue 3, pp 277–294, 2013

\bibitem {BC} Burcnall J. L. Chaundy T. W. {\it A set of differential equations which can be solved by polynomials, Proc. London Soc}, 1929.

\bibitem{CFV1} O.A.Chalykh, M.V.Feigin, A.P.Veselov, {\it New integrable generalizations of Calogero-
Moser quantum problem}, J. Math. Phys., 1998, v.39(2), p.695–703

\bibitem{CFV} O.A. Chalykh, M.V. Feigin, A.P. Veselov, {\it Multidimensional Baker-Akhiezer functions and Huygens' principle}, Commun. Math. Phys., 206 (3), 533-566 (1999)

\bibitem{CSV} Chalykh O.A., Styrkas K.L. and Veselov A.P., {\it Algebraic Integrability for Schrodinger Equations and Finite Reflection Groups}, Theor. Math. Phys., (1993), V.94, N 2, 253–275

\bibitem{C} John Cardy, \emph{SLE for theoretical physicists}, Ann.Phys. 318 (2005) 81-118

\bibitem{CM} L. Carleson and N. Makarov,  {\it Laplacian path models}, J. Analyse Math. 87 (2002), 103-150

\bibitem{Cr} D. Crowdy, {\it Quadrature domains and
fluid dynamics}, In Quadrature domains
and their applications, volume 156 of Oper. Theory Adv. Appl., pages 113-129.
Birkhauser, Basel, 2005.

\bibitem{D} B.Duplantier, \emph{Conformally invariant fractals and potential theory}, Phys.Rev.Lett., 84(7): 1363-1367,(2000)

\bibitem{DNNZ} Bertrand Duplantier, Thi Phuong Chi Nguyen, Thi Thuy Nga Nguyen, Michel Zinsmeister, \emph{The Coefficient Problem and Multifractality of Whole-Plane SLE and LLE}, Annales Henri Poincare,  Volume 16, Issue 6, pp 1311 - 1395, (2015)

\bibitem{G}
Ilya A.~Gruzberg, \emph{Stochastic geometry of critical curves,
Schramm--Loewner evolutions and conformal field theory}, J.~Phys.~A:
Math.~Gen. \textbf{39}, no.~41 (2006) 12601--12655.

\bibitem{GS} B. Gustafsson and H. S. Shapiro, {\it What is a quadrature domain?}, In Quadrature
domains and their applications, volume 156 of Oper. Theory Adv. Appl., pages
1-25. Birkhauser, Basel, 2005.

\bibitem{GTV} B. Gustafsson, R. Teodorescu, and A. Vasil'ev, {\it  Classical and stochastic Laplacian growth}, Advances in Mathematical Fluid Mechanics. Birkhauser/Springer,
Cham, 2014

\bibitem{GMMMO} A. Gerasimov, A. Marshakov, A. Mironov, A. Morozov and A. Orlov, {\it Matrix models of 2D gravity and Toda theory}, Nucl. Phys. B357 (1991), 565-618.

\bibitem{E} P. I. Etingof, {\it Integrability of filtration problems with a moving boundary}, Dokl. Akad. Nauk SSSR, 313(1): 42-47, 1990

\bibitem{EL} A. Eremenko, E. Lundberg, {\it Non-algebraic quadrature domains}, Potential Analysis,38(2013), 787-804

\bibitem{ES} J. Escher and G. Simonett, {\it On Hele-Shaw models with surface tension}, Math.
Res. Lett., 3(4):467-474, 1996.

\bibitem{H} T. Halsey, {\it Diffusion limited aggregation: a model for pattern
formation}, Physcis Today 53 (2000) 36

\bibitem{HJKPS} T.C. Halsey, M.H. Jensen, L.P. Kadanoff, I. Procaccia, and B.I. Shraiman, \emph{Fractal measures and their singularities: the characterization of strange sets}, Phys. Rev. A (3), 33(2): p. 1141, 1986.

\bibitem{Has}
Matthew B.~Hastings, \emph{Exact Multifractal Spectra for Arbitrary
Laplacian Random Walks}, \emph{Phys. Rev. Lett.} \textbf{88} (2002)

\bibitem{HL} M.B. Hastings, L.S. Levitov, {\it Laplacian growth as one-dimensional turbulence}, Physica D, 116 (1998), 244-252

\bibitem{HLO} Sam D. Howison, Igor Loutsenko, John R. Ockendon:
{\it A class of exactly solvable free-boundary inhomogeneous porous medium flows}, Appl. Math. Lett. 20(1): 93-97 (2007)

\bibitem{HLY} J. Harnad, I. Loutsenko, O. Yermolayeva,
{\it Constrained Reductions of 2D dispersionless Toda Hierarchy,
Hamiltonian Structure and Interface Dynamics}, J.Math.Phys. 46 (2005)
112701

\bibitem{JS} Daniel D. Joseph, Jean Claude Saut, {\it Short Wave Instabilities and Ill-Posed Initial Value Problems}, Theoret. Comp. Fluid Dynamics, 1: 191-227 (1990)

\bibitem{K} L. Karp, {\it Construction of quadrature domains in $\mathbb{R}^n$ from quadrature domains in $\mathbb{R}^2$}, Complex Var. Elliptic Eq.,17(1992), 179-188

\bibitem{KL} L. Karp and E. Lundberg, {\it A four-dimensional Neumann ovaloid}, Ark. Mat.,
55(1):185-198, 2017

\bibitem{KKMWZ} I.K. Kostov, I. Krichever, M. Mineev-Weinstein, P.B. Wiegmann, A. Zabrodin, {\it The $\tau$-function for analytic curves}, MSRI Publ., Vol. 40, 285-299 (2001) [Random Matrix Models and Their Applications, Ed. by P.Bleher, A.Its. Cambridge University Press, 2001]

\bibitem{KMWZ} I. Krichever, M. Mineev-Weinstein, P. Wiegmann, A. Zabrodin, {\it Laplacian Growth and Whitham Equations of Soliton Theory}, Physica D198 (2004) 1-28

\bibitem{KMP} D. Khavinson, M. Mineev-Weinstein, and M. Putinar, {\it Planar elliptic growth.
Complex Anal. Oper. Theory}, 3(2):425-451, 2009.

\bibitem{GL}
Gregory F.~Lawler, \emph{Conformally invariant processes in the
plane}, Mathematical Surveys and Monographs, \textbf{114}, Amer.
Math. Soc., Providence, RI, (2005).

\bibitem{GL2}
Gregory F.~Lawler, \emph{Conformal invariance and 2D statistical
physics}, \emph{Bull. Amer. Math. Soc.} \textbf{46} (2009) 35--54.

\bibitem{Lu} E. Lundberg, {\it Laplacian growth, elliptic growth, and singularities of the Schwarz potential}, J. Phys. A: Math. Theor.,44(2011), 135202

\bibitem{L} I. Loutsenko, {\it The variable coefficient Hele-Shaw problem,
Integrability and quadrature identities},
Comm. Math.Phys 268, (2006), no.2, 465-479

\bibitem{LS} Loutsenko I.M., Spiridonov V.P., {\it Self-similar potentials and Ising models}, Pis'ma v ZhETF (JETP Letters) 66 (1997), 747-753

\bibitem{LS_1D} Loutsenko I., Spiridonov V., {\it Spectral self-similarity, one-dimensional Ising chains and random matrices}, Nuclear Phys. B 538 (1999), 731-758

\bibitem{LS_2D} Loutsenko Igor, Spiridonov V, {\it Soliton solutions of integrable hierarchies and Coulomb plasmas} J.Stat.Phys., 99, (2000)

\bibitem{LY1} I. Loutsenko, O. Yermolaeva, {\it Non-Laplacian growth: Exact
results}, Physica D 235, (2007), Pages 56-61

\bibitem{LSLE}
Igor Loutsenko, \emph{SLE$_\kappa$: correlation functions in  the
coefficient problem}, J. Phys. A: Math. Theor. 45 275001,
2012,(2012)

\bibitem{LYSLE} Igor Loutsenko, Oksana Yermolayeva,
\emph{On Exact Multi-Fractal Spectrum of the Whole-Plane SLE}, J.Stat. Mech., (2013) doi: 10.1088/1742-5468/2013/04/P4007

\bibitem{LYLLE} Igor Loutsenko, Oksana Yermolayeva, {\it New exact results in spectra of stochastic Loewner evolution}, J. Phys. A, 47(16):165202, 15, 2014.

\bibitem{LYLLEFP} Igor Loutsenko, Oksana Yermolayeva, {\it Stochastic Loewner Evolutions, Fuchsian Systems and Orthogonal Polynomials}, arXiv:1904.01472

\bibitem{LYZ} Igor Loutsenko, Oksana Yermolayeva, Michel Zinsmeister: {\it On a
competitive model of Laplacian growth}, \emph{J. Statist. Phys.}, 145
(2011) 919--931.

\bibitem{MM} R. McDonald and M. Mineev-Weinstein, {\it Poisson growth}, Analysis and Mathematical Physics, 5(2):193-205, Jun 2015

\bibitem{MPST} J. Mathiesen, I. Procaccia, H. L. Swinney, and M. Thrasher, {\it The universality class of diffusion-limited aggregation and viscous fingering}, Europhysics Letters
(EPL), 76(2):257-263, oct 2006.

\bibitem{MJD} Miwa T., Jimbo M., Date E., Solitons: Differential Equations, Symmetries and Infinite-Dimensional Algebras, Cambridge University Press, 2000

\bibitem{M} Mark Mineev-Weinstein, {\it Selection of the Saffman-Taylor Finger Width in the Absence of Surface Tension: An Exact Result}, Phys. Rev. Lett. 80, 2113 (1998)

\bibitem{MPT} M. Mineev-Weinstein, M. Putinar and R. Teodorescu, {\it Random matrix theory in 2D, Laplacian growth, and operator theory}, J. Phys. A: Math. Theor. 41 (2008)

\bibitem{MWZ} Mark Mineev-Weinstein, Paul B. Wiegmann, Anton
Zabrodin, {\it Integrable Structure of Interface Dynamics}, Phys.Rev.Lett.
84 (2000) 5106-5109

\bibitem{ORGK}
P. Oikonomou, I. Rushkin, I. A. Gruzberg, and L. P. Kadanoff,
{\it Global properties of stochastic Loewner evolution driven
by L\'evy processes}, J. Stat.
Mech. (2008) P01019

\bibitem{ROKG}
I. Rushkin, P. Oikonomou, L. P. Kadanoff, and I. A. Gruzberg, {\it
Stochastic Loewner evolution driven by Levy processes}, J. Stat.
Mech. (2006) P01001

\bibitem{PK} Polubarinova-Kotschina, P.J., {\it On the displacement of the oilbearing contour.},
C. R. (Dokl.) Acad. Sci. URSS, n. Ser. 47. (1945), 250–254

\bibitem{PM} Silvina Ponce Dawson, MarkMineev-Weinstein, {\it Long-time behavior of the N-finger solution of the Laplacian growth equation}, Physica D,
v.73(4), 1994, Pages 373-387

\bibitem {R} Richardson S. {\it Hele Shaw flows with a free boundary produced
by the injection of fluid into a narrow channel}. J.Fluid Mech., 56,
part 4, (1972), pp.609-618

\bibitem{RZ} Steffen Rohde, Michel Zinsmeister, {\it Some remarks on Laplacian growth }, Topology and its Applications,  V(152), Issues 1–2, Pages 26-43, (2005)

\bibitem{S} Selander, G., {\it Two deterministic growth models related to diffusion-limited aggregation}, Thesis. KTH, Stockholm, (1999)

\bibitem{Sh} Shapiro, H. S., {\it The Schwarz Function and Its Generalization to Higher Dimensions}, Arkansas Lecture Notes in the Mathematical Sciences 9, John Wily \& Sons, Inc., New York, 1992

\bibitem{ST} Saffman, P. G., Taylor, G, {\it The Penetration of a Fluid into a Porous Medium or Hele-Shaw Cell Containing a More Viscous Liquid}, Proceedings of the Royal Society A: Mathematical, Physical and Engineering Sciences, 245(1242), 312–329, (1958)

\bibitem{VE} A.N. Varchenko and P.I. Etingof, {\it Why the boundary of a round drop becomes
a curve of order four},
American Mathematical Society, University Lecture Series, 3, (1994)

\bibitem{Z} A.Zabrodin, {\it New applications of non-hermitian random matrices}, Annales Henri Poincare 4 (2003) S851-S861

\bibitem{Z1} A.Zabrodin, {\it Canonical and grand canonical partition functions of Dyson gases as tau-functions of integrable hierarchies and their fermionic realization}, Complex Analysis and Operator Theory, 4 (2010) 497-514

\bibitem{Z2} A.Zabrodin, {\it Growth of fat slits and dispersionless KP hierarchy}, J.Phys.A: Math. Theor., 42 (2009) 497-514

\bibitem{HHD} Halsey, T. C., Honda, K., Duplantier, B. (1996), {\it Multifractal dimensions for branched growth}, J.Stat.Phys., 85(5-6), pp 681–743

\bibitem{Wiki} \url{http://en.wikipedia.org/wiki/Multifractal_system}

\end{thebibliography}
\end{document}